\documentclass[usegraphicx]{mn2e}

\usepackage{fixltx2e}
\usepackage{mathptmx}

\newcommand{\apj}{ApJ}
\newcommand{\mnras}{MNRAS}
\newcommand{\nat}{Nat}

\newcommand{\physrevD}{Phys. Rev. D}
\newcommand{\aap}{A\&A}                   
\newcommand{\aaps}{A\&AS}                 
\newcommand{\aj}{AJ}                      
\newcommand{\pasp}{PASP}                  
\newcommand{\apjl}{ApJ}                   

\begin{document}

\title[Diffuse Coma]{Diffuse Radio Emission in/around the Coma Cluster: Beyond Simple Accretion}

\author[Brown \& Rudnick]{Shea Brown$^{1,2}$ \& 
Lawrence Rudnick$^{3}$ \\ $^{1}$CSIRO, Australia Telescope National 
Facility, P.O. Box 76, Epping, NSW 1710, Australia\\ $^{2}$Bolton Fellow \\ $^{3}$Department of 
Astronomy, University of Minnesota, Minneapolis, MN 55455} \maketitle

\begin{abstract} We report on new 1.41~GHz Green Bank Telescope and 352~MHz Westerbork Synthesis Radio Telescope observations of the Coma cluster and its environs. At 1.41~GHz we tentatively 
 detect an extension to the Coma cluster radio relic source 1253+275 which makes its total extent $\sim$2~Mpc. This extended relic is linearly polarized as seen in our GBT data, the NVSS, and archival images, strengthening a shock interpretation. The extended relic borders a previously undetected ``wall" of galaxies in the infall region of the Coma cluster. We suggest that the radio relic is an infall shock, as opposed to the outgoing merger shocks believed responsible for other radio relics.  We also find a sharp edge, or ``front", on the western side of the 352~MHz radio halo. This front is coincident with a similar discontinuity in the X-ray surface brightness and temperature in its southern half, suggesting a primary shock-acceleration origin for the local synchrotron emitting electrons. The northern half of the synchrotron front is less well correlated with the X-ray properties, perhaps due to projection effects. 
We confirm the global pixel-to-pixel power-law correlation between the 352~MHz radio brightness and X-ray brightness with a slope that is inconsistent with predictions of either primary shock acceleration or secondary production of relativistic electrons in Giant Radio Halos, but is allowable in the framework of the turbulent re-acceleration of relic plasma.  The failure of these first order models and the need for a more comprehensive view of the intracluster medium energization is also highlighted by the very different shapes of the diffuse radio and X-ray emission. 
We note the puzzling correspondence between the shape of the brighter regions of the radio halo and the surface mass density derived from weak lensing.  
 \end{abstract}

\begin{keywords}
  galaxies: clusters --- intergalactic medium --- acceleration of particles --- shock waves ---
  magnetic fields --- galaxies: clusters: individual: Coma --- intergalactic medium --- dark matter 
  --- X-rays: galaxies: clusters
\end{keywords}

\section{Introduction} 

Observations of diffuse radio emission in clusters of galaxies currently 
provide our only window into the relativistic particle population that 
potentially makes up a non-trivial fraction of the energy density in the 
Intra-Cluster Medium (ICM) (e.g., Skillman et al. 2008). Large-scale 
features such as Giant Radio Halos (GRHs) and peripheral relics have 
been found in more than 50 clusters to date (Ferrari et al. 2008). 
Though the origin of the seed relativistic electrons is still unknown, 
all of these systems appear to have recently undergone at least some 
minor mergers/accretion. This highlights the potential to use diffuse 
synchrotron emission to illuminate ICM
energization in both clusters and lower density regions invisible at 
other wavelengths (e.g., Pfrommer et al. 2007; Rudnick et al. 2009).


There are three broad potential sources for the cosmic-ray electrons (CRe) in the ICM: extended radio galaxies (AGN models), electrons directly accelerated by accretion/merger shocks and/or cluster turbulence (primary models; En{\ss}lin et al. 1998; Brunetti et al. 2001), and electrons created from relativistic proton-proton collisions from cosmic-ray protons (CRp) that have accumulated in the cluster's potential well (secondary models; Dennison 1980). The two favoured  mechanisms for producing {\it spatially distributed} CRe is the turbulent re-acceleration of mildly relativistic electrons, and broadly distributed CRp resulting in secondary leptons. Observations of a large sample of massive clusters currently favours the turbulent re-acceleration model (Venturi et al. 2007, 2008; Brunetti et al. 2009).  A ``unified model" for GRHs and radio relics (Pfrommer et al. 2008) using both primary shock acceleration and secondary models has also been proposed, and will be tested by the next generation of low-frequency radio and gamma-ray telescopes. 

The Coma cluster is a classic example of a merging/accreting system that 
hosts both a GRH (Coma C; Willson 1970) and peripheral radio relic 
(1253+275; Jaffe \& Rudnick 1979; Giovannini et al. 1985) and has 
been used as a test-bed for CR acceleration models (e.g., Donnert et al. 
2009). The dynamical state of Coma has been well established through 
optical velocity analysis (Fitchett \& Webster 1987; Mellier et al. 
1988; Merritt \& Trimbley 1994; Colless \& Dunn 1996; and Adami et al. 
2005).

Coma has been extensively observed in the radio. Kim et al. (1989) and 
Venturi et al. (1990) mapped the region with the Westerbork Synthesis 
Radio Telescope (WSRT) at 326~MHz, finding a diffuse ``bridge" of 
emission connecting the halo to the relic. Deiss et al. (1997) also 
detected the bridge at 1.4 GHz using the Effelsberg 100-m-telescope, and 
found the radio halo to have a diameter of $\sim$80$^{\prime}$. Using 
the 300~m Arecibo telescope combined with DRAO data at 408~MHz Kronberg 
et al. (2007) detected a $\sim$135$^{\prime}$ radio ``cloud" surrounding 
both the classical halo and relic source. This cloud size corresponds to 
$\sim$4~Mpc if it is associated with the Coma cluster, and would 
therefore extend into the Warm-Hot Intergalactic Medium (WHIM). The Coma 
cluster is also one of the few GRHs to have resolved spectral-index 
maps (Giovannini et al. 2003), which show a spectral steepening at 
larger cluster radii. Recent $\sim$150~MHz observations with the 
Westerbork Synthesis Radio Telescope (Pizzo 2010) confirm the 
radial steepening of the spectral index, as well as identifying two new 
candidate relic features to the East and West of the cluster.

Coma has also been the subject of intense X-ray observations. In 
addition to surveys (e.g. the ROSAT All Sky Survey), mosaic $XMM-Newton$ 
observations (Briel et al 2001; Neumann et al. 2003; Schuecker et al. 
2004)   have revealed the complex thermal substructure of the X-ray halo. 
Detections of non-thermal hard X-ray emission have been claimed, e.g., 
using $RXTE$ (Rephaeli \& Gruber 2002) and $BeppoSAX$ (Fusco-Femiano et 
al. 1999, 2004). Diffuse {\it thermal} hard X-ray emission has been 
imaged with $INTEGRAL$ (Renaud et al. 2006; Eckert et al. 2007; 
Lutovinov et al. 2008) and $Suzaku$ (Wik et al. 2009), revealing 
temperature increases in the Western region of the cluster also seen at lower
energies with $ASCA$ (Watanabe et al. 1999).

We present Green Bank Telescope (GBT) 1.41~GHz and (WSRT) 352~MHz 
observations of the Coma cluster in an attempt to investigate the 
various CR acceleration models, as well as confirm the dramatic 4~Mpc 
radio cloud seen by Kronberg et al. (2007). We present the observations 
and data reduction in $\S$2, and our observational results in $\S$3. In 
$\S$4 we present a discussion of the implications of our results, and 
summarize our key points in $\S$5.

In this paper, we assume $H_{o}=70$, $\Omega_{\Lambda}=0.7$, 
$\Omega_{M}=0.3$.

\begin{figure*}
\includegraphics[width=16cm]{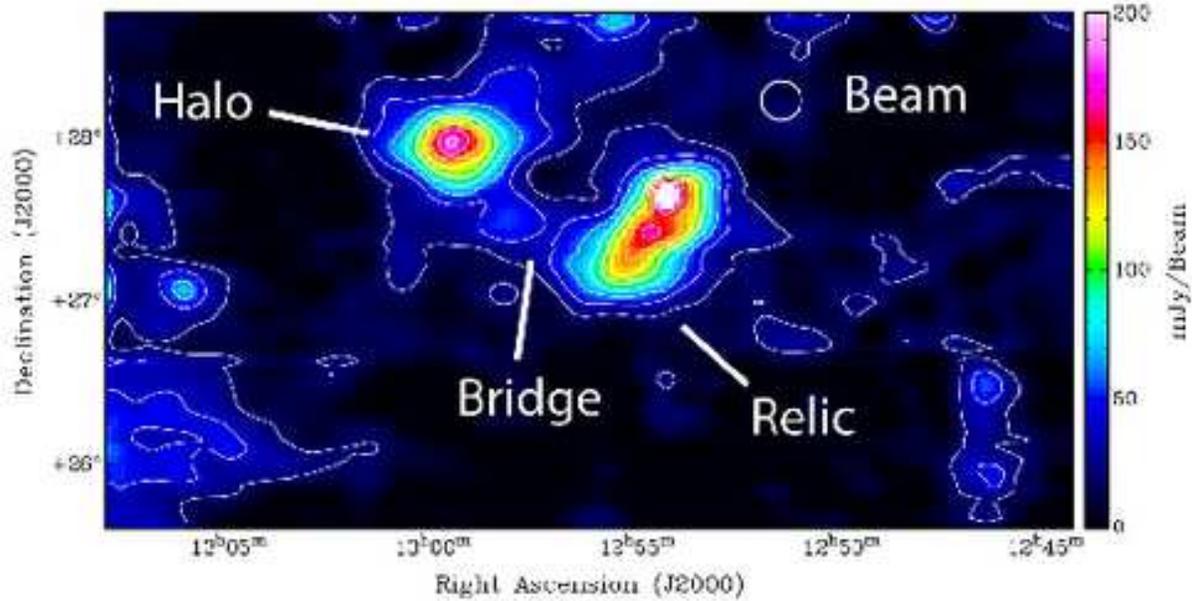}

\caption{\label{gbt}GBT total intensity image (using DEC scans only) with all NVSS emission
subtracted out. (14.25$^{\prime}\times$13$^{\prime}$ beam). Contours start at 20 mJy~beam$^{-1}$ (3$\sigma$) and increase in steps of 20 mJy~beam$^{-1}$. The bright background source Coma A (3C 277.3) was fit and subtracted out by hand, although a small residual is still seen as the burnt out spot towards the northern part of the relic.}

\end{figure*}

\section{Observations \& Data Reduction}
 We observed a 6$^{\circ}$x6$^{\circ}$ region around the Coma Cluster 
with the Green Bank Telescope (GBT) between October 2007 and January 
2008. We observed with the GBT's Spectrometer with a 50 MHz bandpass 
centred on 1.41 GHz. The region was sampled with 150 evenly spaced 
stripes -- 75 with constant RA and 75 with constant Dec. Each 6$^{\circ}$ 
stripe was covered by five 60 second raster scans. We performed these 
scans immediately one after another in order to identify and edit out 
those times when the overall power levels were unstable due to either 
receiver or atmospheric fluctuations. During each night of observations, 
we scanned across some combination of 3C295, 3C48, 3C138 and 3C286. The 
first two are unpolarized calibrators and the second two are polarized. 
There were some nights where, for a variety of reasons, we did not 
observe an unpolarized calibrator. An internal correlated calibrator 
signal ($\sim$19~K) was used to determine the relative X and Y dipole 
gains (correcting for leakage from Stokes I into Stokes Q), as well as 
the X-Y phase offset. A parallactic angle correction, which rotates 
Q$_{Telescope}$ and U$_{Telescope}$ into Q$_{Sky}$ and U$_{Sky}$, was 
applied to each integration. Observations of the polarized calibrators 
over 3 independent parallactic angles allows, in principle, for the determination of the 
full Mueller matrix elements (e.g., Mason 2007), which describe the conversion 
of measured Stokes parameters into true source Stokes parameters. We 
were unable to determine these elements due to insufficient parallactic 
angle coverage of the polarized calibrators. This results in the loss of 
$<$1\% of the U Stokes power through conversion into V, based on 
fractional polarization measurements of 3C138 \& 3C286.

\begin{figure}
\includegraphics[width=8cm]{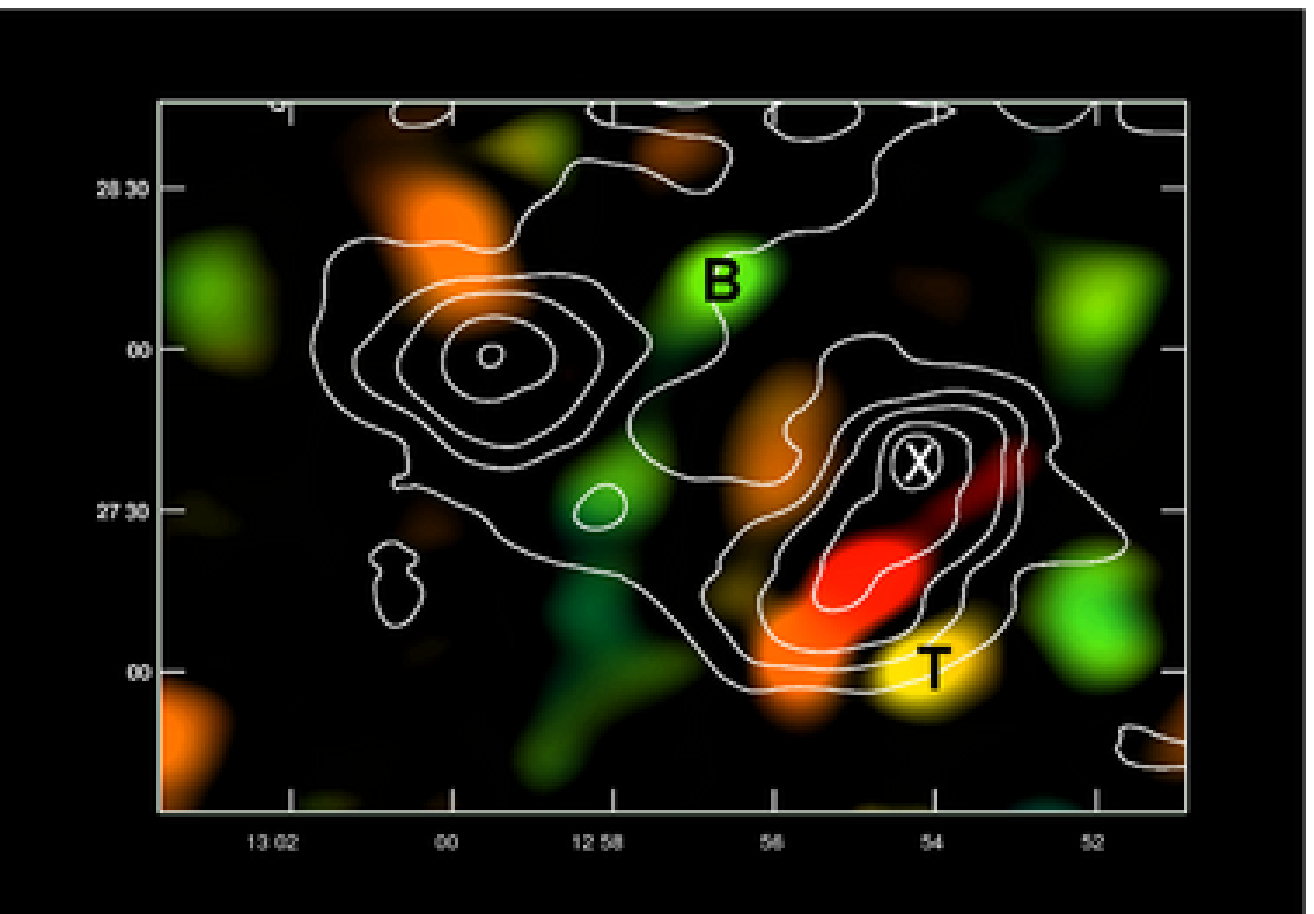}
\includegraphics[width=8cm]{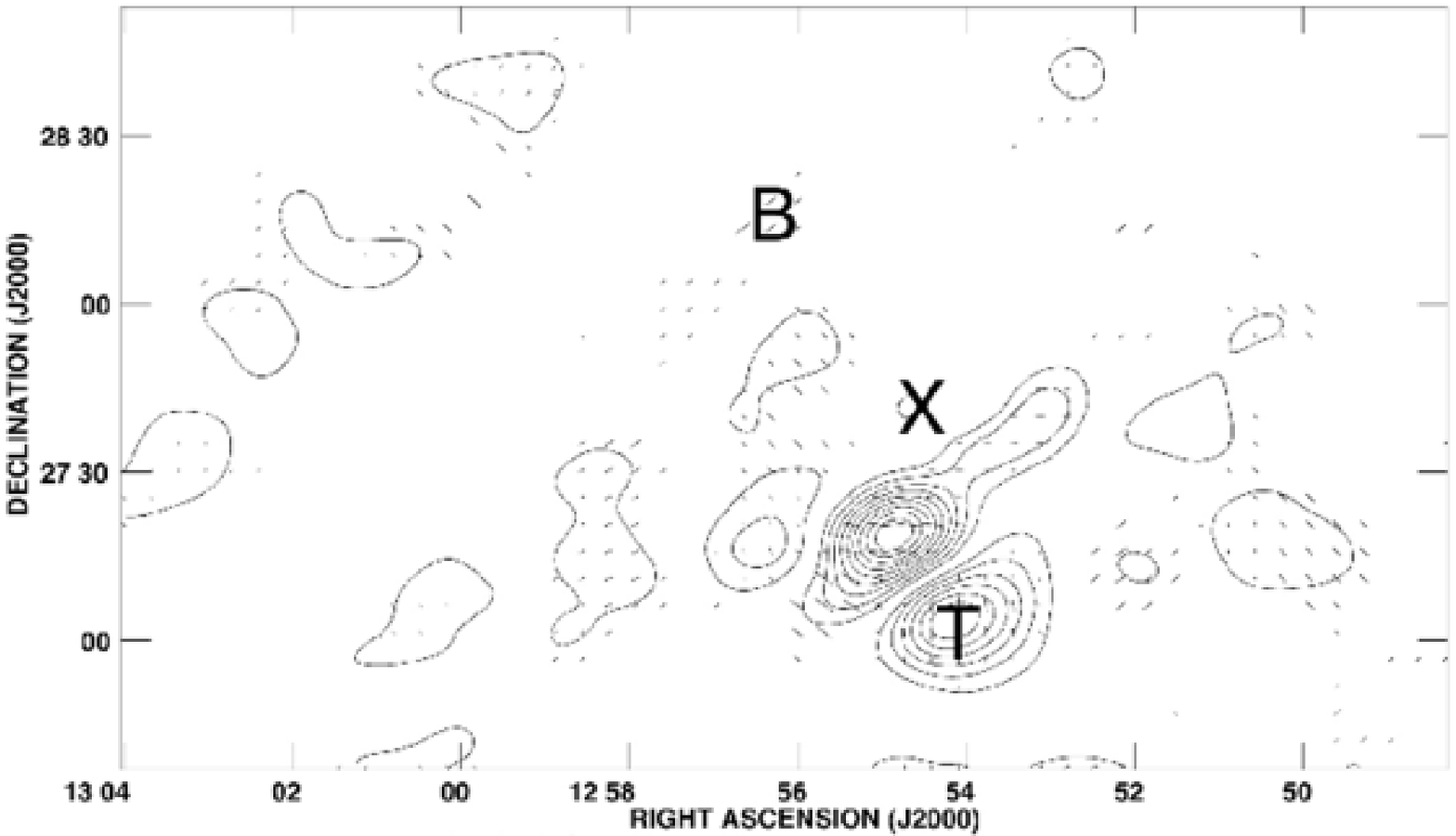}
\includegraphics[width=8cm]{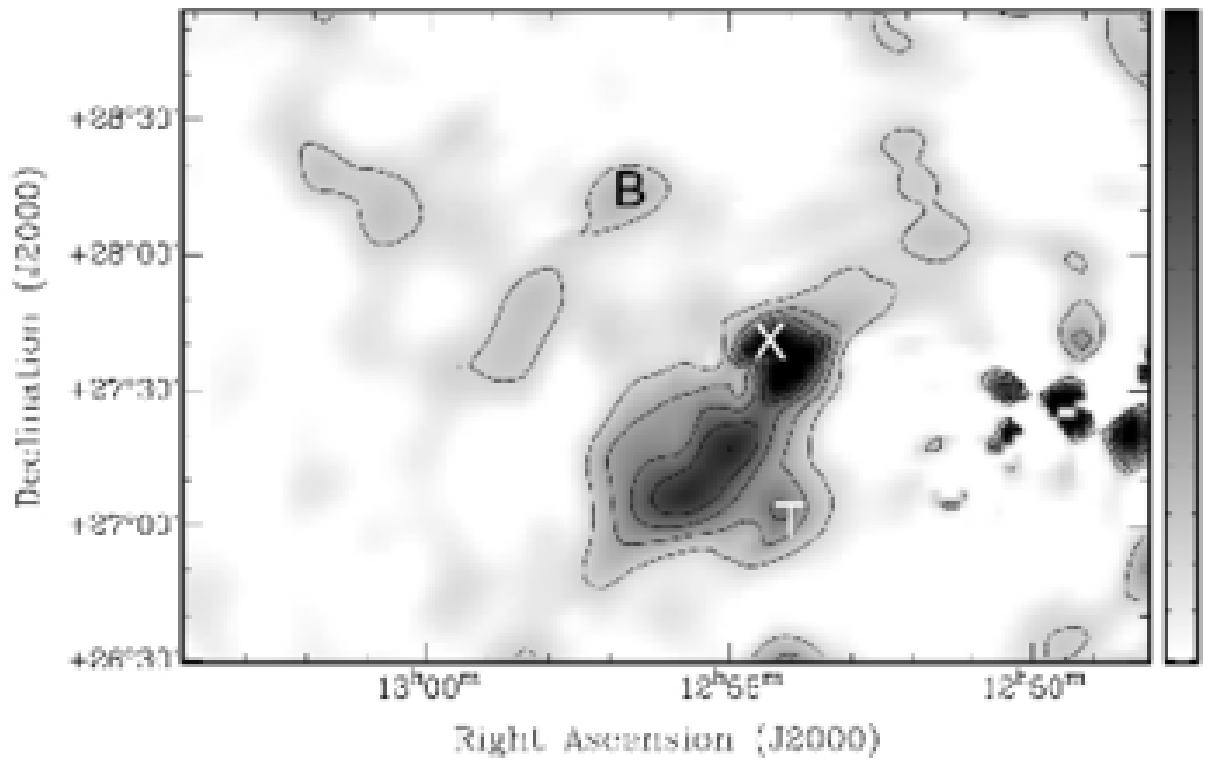}

\caption{\label{gbt_pol} Top: Contours are GBT total intensity (Fig. 
\ref{gbt}) over polarized intensity colour coded by polarization angle. Both images have a beam of 898$^{\prime\prime}\times$876$^{\prime\prime}$ at 106$^{\circ}$ position angle. The bright background source Coma A (3C 277.3) was fit and subtracted out by hand (located at the X). Contour levels of total intensity are at 13.24$\times$(-1,1, 2, 3, 5, 7, 10) mJy~beam$^{-1}$. The polarized intensity varies from 0 to 25.6 mJy~beam$^{-1}$. The three diffuse polarized patches that make up the extended relic have polarization angles of (south to north) 55$^{\circ}$, 80$^{\circ}$, and 94$^{\circ}$ , and sources T and B have angles -0.7$^{\circ}$ and -47$^{\circ}$ respectively. Centre: Absolute value of Stokes $Q$ contours with polarization vectors over-plotted. Contours are 2.27E-03$\times$(1,2,3,4,5,6,7,8,9,10) Jy~beam$^{-1}$. 
 1$^{\prime\prime}=55~\mu$Jy~beam$^{-1}$; Bottom: Contours are polarized intensity from the NVSS, convolved to 800$^{\prime \prime}$ (Rudnick \& Brown 2009), starting at 3$\sigma$=4.8~mJy~beam$^{-1}$ and increasing in 3$\sigma$ intervals. The colour scale is linear running from 0 (white) to 27~mJy~beam$^{-1}$ (black).}

\end{figure}

Due to radio interference and instabilities in the Spectrometer gains, $\sim$35\% of the data was unusable. For this paper, we only used the stripes taken at constant declinations, which were more stable than the constant RA stripes.  Fortunately, most of the bad data was in the outer regions of the image.  In order to isolate the diffuse emission, we subtracted out a convolved version of the corresponding NVSS images.  A linear baseline was then removed from each scan by fitting a straight line to source free regions after the NVSS subtraction. The resulting map rms is $\sim$6~mJy~beam$^{-1}$ (which is dominated by diffuse Galactic emission), where the beam is 14.25$^{\prime}\times$13$^{\prime}$. Fig. \ref{gbt} shows the total intensity map after NVSS subtraction and baseline removal.  The low resolution comes from our rapid on-the-fly mapping of the region, needed in order to mitigate gain fluctuations. Figure \ref{gbt_pol} shows the polarized intensity colour coded by polarization angle with total intensity contours. We describe the halo and relic emission in $\S$3. In both figures, the radio source Coma A (3C277.3) was fit with a gaussian and subtracted out of the image by hand.

 A four-pointing mosaic (pointing centres 12h59m52s+27d58m, 12h54m08s+27d58m, 12h59m52s+26d42m, and 12h54m08s+26d42m) of the Coma cluster was also observed for a total of $\sim$48 hours over four nights in P-band (352~MHz) with the 
Westerbork Synthesis Radio Telescope (WSRT) in November of 2008. The array was in the maxi-short configuration, with the shortest baselines of 36m, 54m 72m and 90m, used to optimise imaging of very extended structures. One primary flux and one polarized calibrator (as a pair) were observed at the beginning and end of each night.  For the primary flux 
calibrators we observed 3C147 and 3C295, and for the polarized calibrators we observed DA240 and 3C345. We used the WSRT wide band correlator to cover a frequency range from 310-390~MHz with eight 10~MHz wide bands, each with 128 channels and full Stokes parameters. IFs 3, 4, 6, and 8 were not used due to interference and calibration problems. After 
removing the end channels in each band and editing for strong RFI, 400 channels remained in the final analysis, for a total bandwidth of 31~MHz.

 The calibration and reduction of the WSRT data were performed using the NRAO's Astronomical Image Processing System (AIPS). The total intensity in each of the 4 bands was calibrated independently using standard procedures and the fluxes in the VLA calibrator manual for 3C147 and 3C295.  We did several iterations of amplitude and phase self-calibration on each data set. Fig. \ref{wsrt_mos} shows the total intensity image which is a combination of the 4 IF images with an average frequency of 352~MHz. In order to see the diffuse emission more clearly, we imaged the point sources by using only UV data $>$ 700~$\lambda$, then subtracted their clean components out of the original UV data and reimaged the residuals at a resolution of $4^{\prime}\times2^{\prime}$, matching that of Kim et al. (1989). Fig. \ref{wsrt_mos} shows contours of the full-resolution 134$^{\prime\prime}\times$68$^{\prime\prime}$ image prior to point-source subtraction, plotted over the low-resolution point-source subtracted image. 

\section{Results} 

\subsection{Extended Relic} The relic emission shown in Fig. \ref{gbt} is much larger and more diffuse than has been previously seen. Giovannini et al. (1991) measured a total extent for the relic of 25$^{\prime}$ using the WSRT and VLA.  Our subtraction of Coma A by hand, and the residuals that are left behind, makes a clean interpretation of Fig. \ref{gbt} difficult. Based on the quality of the subtraction for other point-sources in the image, which show at most a 10\% residual extending up to 9$^{\prime}$ from the peak of emission,  real extended emission appears not only to the South, but also to the West and slightly North of Coma A. However, based {\it solely} on this total intensity image, we characterize the extended relic as a tentative detection. The diffuse relic as seen by the GBT extends through Coma A into the NW, with a total extent of 67$^{\prime}$, or 2~Mpc at the redshift of the Coma cluster. The 408~MHz image of Kronberg et al. (2007) also shows emission extending this far (and farther) to the NW of the known relic, but the halo and relic are not cleanly separated at those frequencies. The centre of the extended relic is $\sim$75$^{\prime}$ (2.2~Mpc) from the X-ray cluster centre.

We also detect the extended relic in linear polarization in the 1.4~GHz GBT measurements (Fig. \ref{gbt_pol}, top), with a fractional polarization between 12-17\%. The polarization angle (PA) varies monotonically from South to North by $\sim$40$^{\circ}$, separated into three distinct patches. This would be inconsistent with a single shock structure if the angles were intrinsic to the source.  This variation in polarization angle, however, is typical of that seen in the polarized Galactic background over similar scales (Tucci et al. 2002) and is likely due to foreground Faraday modulation. There is also the additional complication from the Coma A subtraction, which we performed by hand in the Q and U images. We also detect the polarized emission from the tailed radio galaxy NGC4789 as the bright yellow patch (T) at the southwest edge of the relic.  The bright green patch of polarized emission (B) at 12h56m37s, 28d16' is associated with a distant radio galaxy mapped by Rogora et al. (1986) and also visible in FIRST (Becker et al. 1995).  There is no counterpart to the nucleus visible in the DSS, although the source was apparently misidentified by Fanti et al. (1975) with an optical object on the eastern lobe.  The other patches of polarized flux in Fig. \ref{gbt_pol} do not appear to be associated with either galactic or extragalactic total intensity structures. Fortunately, almost all of the polarized flux from Coma A is in the Stokes U component of our GBT image. We therefore plot the absolute value of Stokes Q contours {\it without Coma A subtraction} in Fig. \ref{gbt_pol} (middle) with polarization vectors (0.5~tan$^{-1}(\frac{U}{Q})$). Again, the polarized emission extends up-to and past Coma A. 

We also sought further confirmation of this extended relic using independent data.  Rudnick \& Brown (2009) produced all-sky maps of polarized intensity from the NVSS by convolving $P=\sqrt{Q^2 + U^2}$ to 800$^{\prime\prime}$, roughly the same resolution as our GBT images. Fig. \ref{gbt_pol} shows the 800$^{\prime\prime}$ NVSS polarization image in the field around the relic. The extended relic, to the west and north of Coma A, as well as emission from Coma A, is clearly present in the NVSS, and extends even farther North at lower levels. In addition, the 11~cm single dish polarized image of Andernach et al. (1984; Fig. 2) shows emission extending from the known relic 1253+275 and passing just West and then North of Coma A, in agreement with Fig. \ref{gbt_pol}. 

The total flux density of the radio relic is highly uncertain due to the patchy galactic emission evident in the GBT images and the confusion from Coma~A; it is between 200-500~mJy at 1.41~GHz. Giovannini et al. (2001) measured only 160~mJy at 1.4~GHz with the VLA, though they are likely missing a significant amount of flux. They find a best fit power-law to the data at other frequencies that predicts $\ge$200~mJy at 1.4~GHz. 

The relic, as seen in our 352~MHz image, is only $\sim$28$^{\prime}$ long before it becomes confused by Coma~A. It has a flux density of 1700~mJy, comparable to the 326~MHz value of 1400~mJy given by Giovannini et al. (1991).  There are several reasons why the extended relic is seen in the GBT 1.4 GHz images and not clearly in the WSRT 350 MHz images. The most important is the instrumental contamination from Coma~A, roughly indicated in Fig. \ref{wsrt_mos}. Also, as seen in the 1.4 GHz GBT image, the emission to the North of 1253+275 is very diffuse, and likely does not have the fine-scale structure of 1253+275. This does not preclude it from being part of the same structure, but 1253+275 is in a region with slightly higher optical and X-ray density (see $\S$4.1), indicating that it may have more mass flux through the shock. This would lead to an increased synchrotron surface-brightness. The presence of ``relic" relativistic plasma could also cause increased surface-brightness for the southern part of the relic. Similar partial illumination of larger shock structures as been proposed for Abell 548 (Solovyeva et al. 2008). Deep interferometric observations with excellent uv-coverage and sensitivity to large-scale emission are needed to confirm these results. We should note, however, that Kronberg et al. (2007) reported that all of the large-scale features seen in the combined Arecibo + DRAO images, which include diffuse emission corresponding to our extended relic, can be seen in the interferometric DRAO images alone.

\subsection{Halo Profile} 
\begin{figure*} 
\includegraphics[width=16cm]{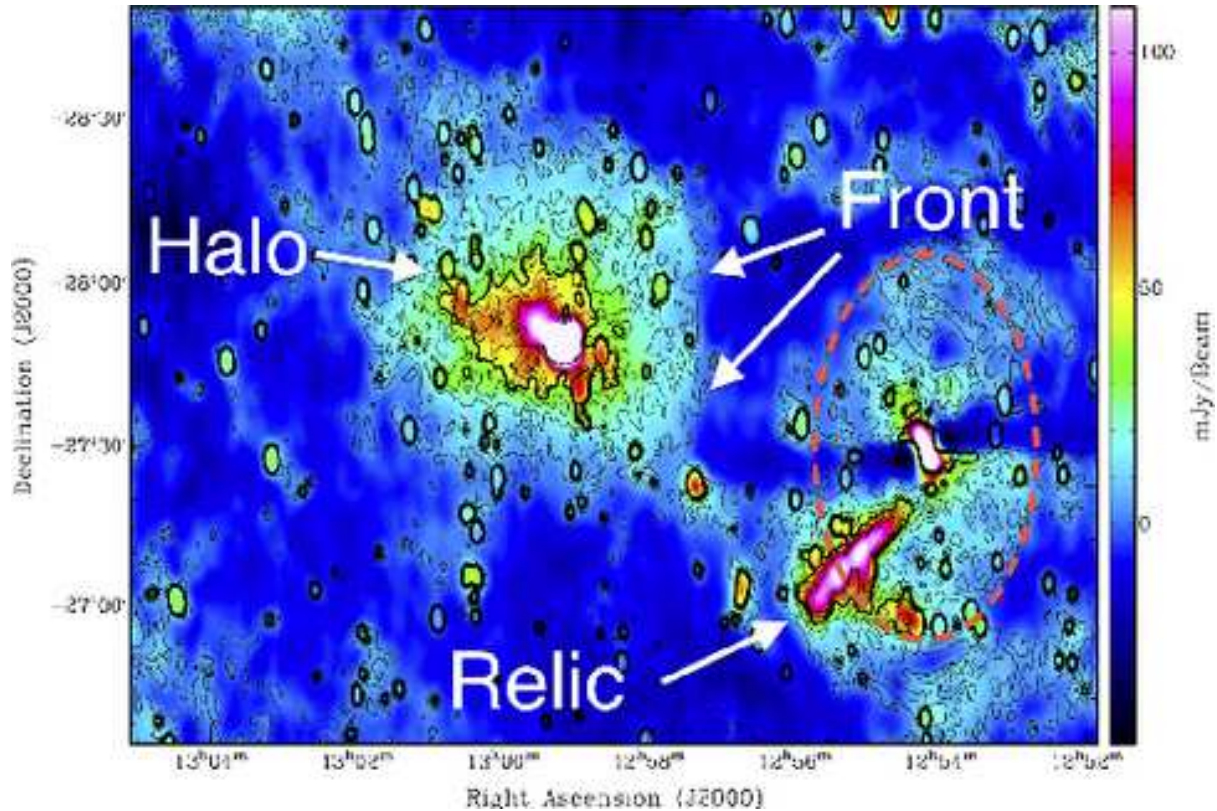} 

\caption{\label{wsrt_mos} WSRT total intensity image of the Coma Cluster. The image was made from a 4-pointing mosaic with a central frequency of 352~MHz (31~MHz bandwidth) and a resolution of 134$^{\prime\prime}\times$68$^{\prime\prime}$. Contours start at 1~mJy~beam$^{-1}$ and increase in intervals of 3~mJy~beam$^{-1}$.  For the colour image we have subtracted the majority of the point-source flux and convolved to $4^{\prime}\times 2^{\prime}$ (see text). The dashed oval indicates the region with spurious emission due to incomplete modelling and subtraction of the background source Coma~A.} 

\end{figure*}

\begin{figure}
\includegraphics[width=8cm]{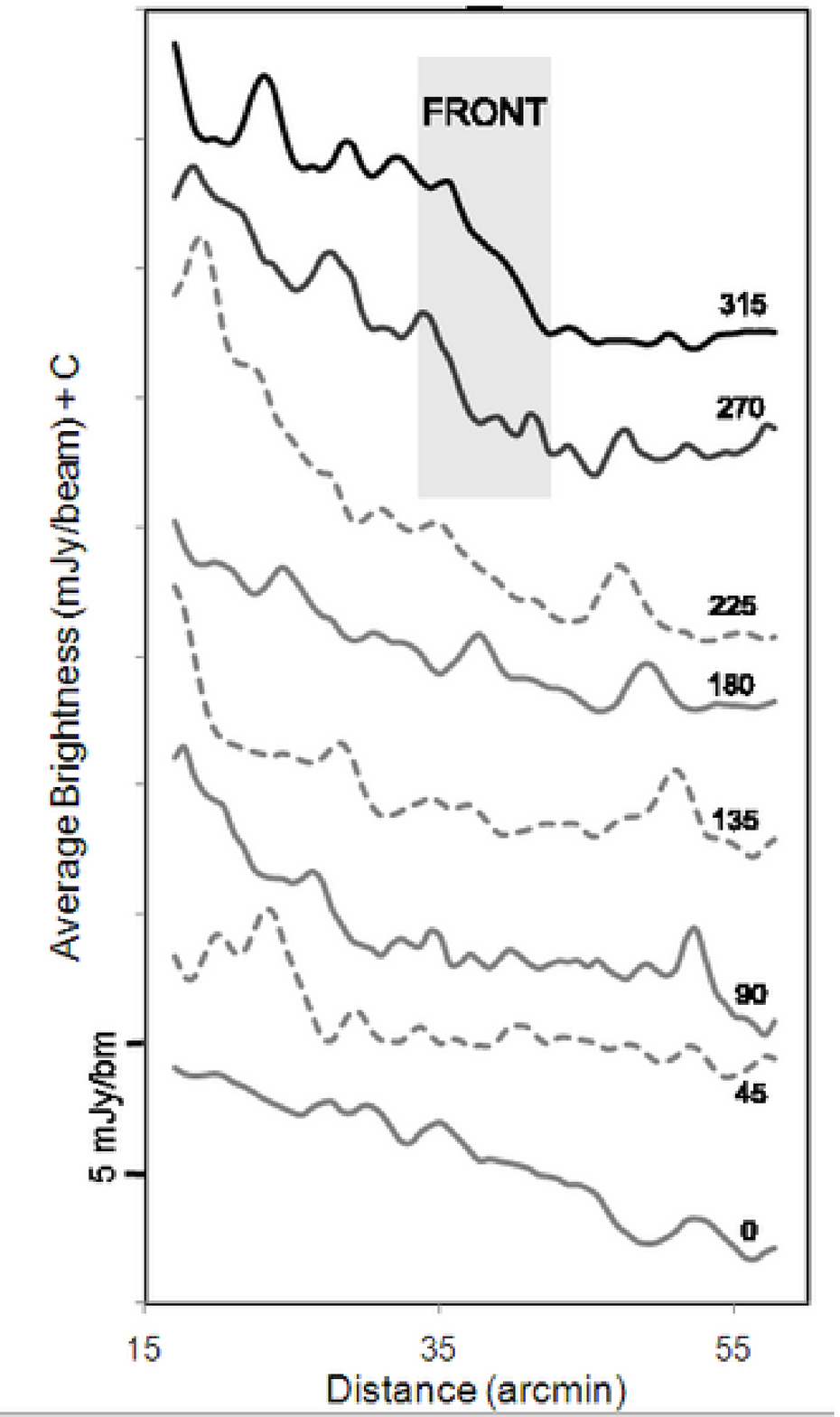}
\caption{\label{annular}  45$^{\circ}$ annular averages of the point-source subtracted WSRT image, at the position angles shown and offset by a constant. N (0$^{\circ}$), E (90$^{\circ}$), S (180$^{\circ}$), and W (270$^{\circ}$).  The Western region has a sharper drop in brightness. relative to the other directions. In the Southwest, emission from the bridge will smear out any sharper features.}

\end{figure}

As indicated in Fig. \ref{wsrt_mos} and \ref{annular}, the Western edge of the halo shows a sharp ``front" or drop in radio surface-brightness. Fig. \ref{annular} shows a plot of azimuthally averaged surface-brightness over 45$^{\circ}$ wedges centred on 13 00 +28 30, at position angles as indicated. The brightness shows a $\sim$5~mJy/beam increase at the front. This feature is located on only one side of the cluster and is unique for GRHs.  A linear structure coincident with the southern half of the western front is also seen in the $\lambda$~=~2m images of Pizzo (2010), although it is not connected to the rest of the halo.   Deiss et al. (1997) reported a 360$^{\circ}$ annular average of the 1.4~GHz radio halo that shows a large departure from the King model at a similar distance as the Western front. Based on Fig. \ref{annular}, the Western front could be a major contributor to this effect. We discuss the implications of this front, as well as confirming multi-wavelength data, in $\S$4.2.

\subsection{Other Diffuse Emission} In our GBT image, the halo has a diameter of roughly 84$^{\prime}$. We confirm the ``bridge" of emission (Kim et al. 1989; Venturi et al. 1990) connecting the halo and relic in both the GBT and WSRT images. To compare with the Venturi et al. (1990) WSRT observations, we convolved Fig. \ref{wsrt_mos} to 4$^{\prime} \times$3$^{\prime}$ resolution and found a bridge surface-brightness of $\sim$25-30~mJy/beam in a brighter region, which is consistent with the Venturi et al. (1990)  4$^{\prime} \times$3$^{\prime}$ resolution image in the same region. We also confirm much of the $\sim$135$^{\prime}$ ``radio cloud" emission (which appears as an low surface-brightness envelope in Fig. \ref{cloud}) first seen by Kronberg et al. (2007) with a total 408~MHz flux density of 0.8-3~Jy.  However, we note that in our GBT image, the envelope (cloud) has a brightness of only $\sim$10-25~mJy/(14.25$^{\prime}\times$13$^{\prime}$ beam). There are fluctuations in the galactic foreground with similar angular scales and brightness in the image (see Fig. \ref{gbt}).

\section{X-ray/Optical Comparisons and Discussion} 

\subsection{Relic}

\begin{figure}
\includegraphics[width=8cm]{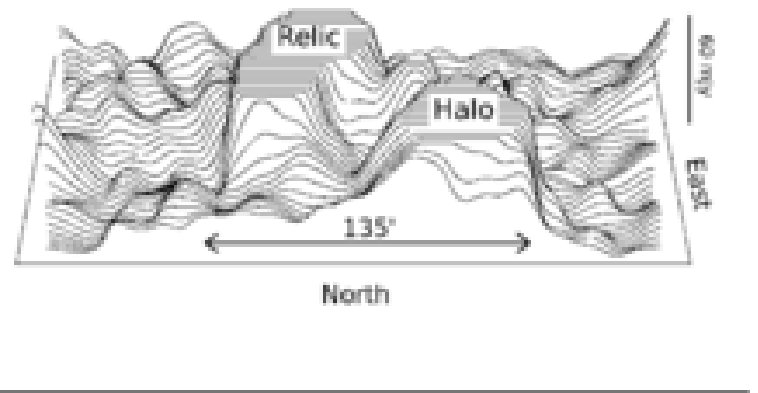}

\caption{\label{cloud} Topographic map of our 1.4~GHz GBT image showing the 135$^{\prime}$ ``radio cloud" first observed by Kronberg et al. (2007). The view is looking from the North. }

\end{figure}
We re-examine the properties of the radio relic and its relationship to the optical and X-ray tracers in the Coma cluster outskirts in light of our finding that it is $\sim$2~Mpc in extent, twice that previously recognised.  The increase in the relic's size and luminosity itself poses no fundamental theoretical problems, nor does it make this an extreme relic observationally (see, e.g., Brown \& Rudnick 2009 for a compilation of observed relics and their radio luminosities L$_{Radio}$ using Giovannini et al. 1999 and other sources).
In fact, the bright, previously identified piece of the relic (1253+275) is on the lower radio luminosity envelope of L$_{Radio}$ vs. L$_{X}$ correlation, as seen in Fig. 15 of Brown \& Rudnick (2009), which has an order of magnitude scatter in L$_{Radio}$ values. 

\begin{figure}
\includegraphics[width=8cm]{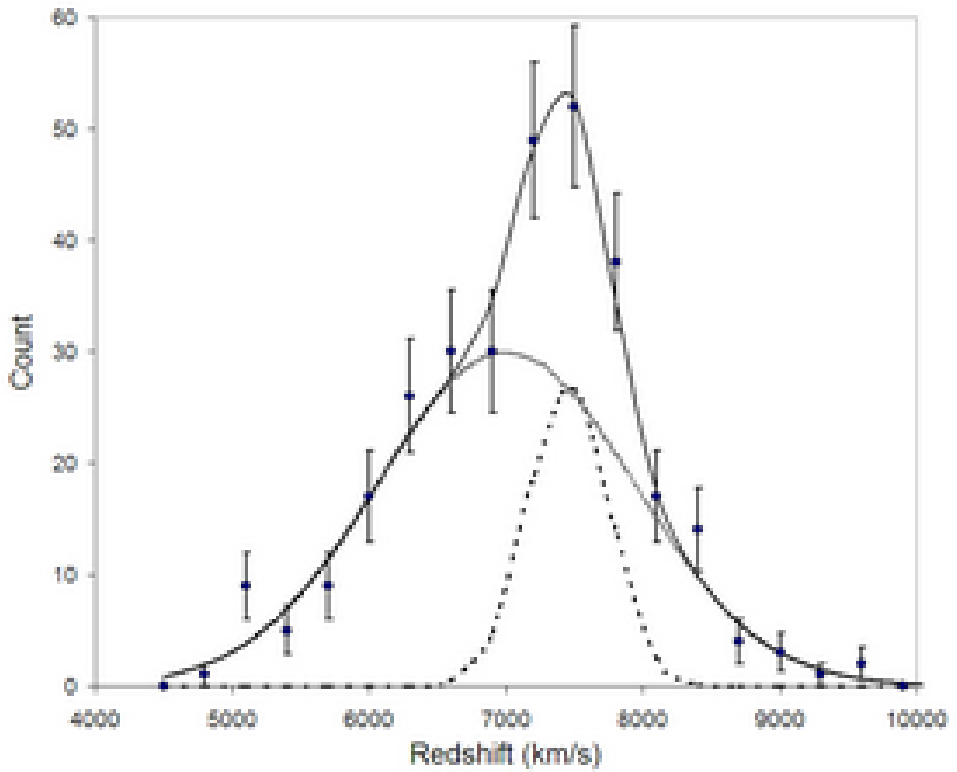}
\includegraphics[width=8cm]{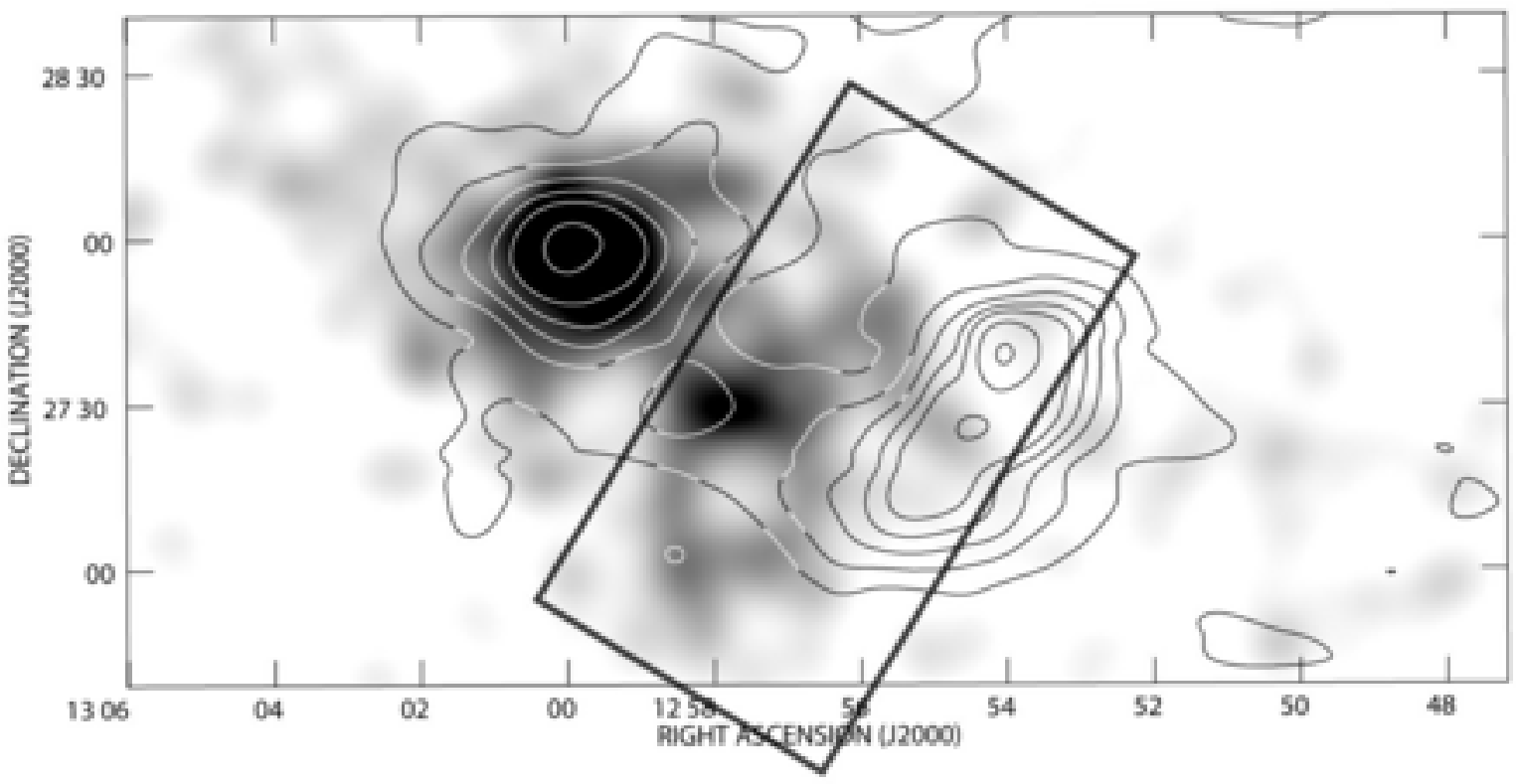}

\caption{\label{vel_plot} Top: Plot of velocities in a 2$^{\circ}$ high
band ranging from 0.5 to 1.5 degrees from the Coma center, shown at
bottom. Bottom: Smoothed surface-density of optical galaxies from the
SDSS spectroscopic database with 6600 $<$ v $<$ 8200 (greyscale) with
GBT contours from Fig 1. Note the ``wall" of galaxies that ends on the
inside edge of the radio relic.}

\end{figure}
Large shock-fronts of this size have been seen for outwardly propagating shocks (R$\ddot{o}$ttgering et al. 1997; Johnston-Hollitt et al. 2002;  Bonafede et al. 2009), but those are spherical while 1253+275 with the newly discovered extension exhibits only mild curvature compared to the size of the cluster. The outgoing shock model also runs into problems with the lack of a temperature rise in the X-ray gas at the position of the relic (Feretti \& Neumann 2006).  They suggest, instead, that the radio emitting particles are energized by turbulence generated by the infall of the NGC~4839 group into a pre-existing relativistic plasma.  An alternative is that the relic represents an ``infall" shock
shock\footnote{  Here we distinguish between what we are calling an ``infall" shock from  ``accretion" and ``merger" shocks. In numerical simulations of large-scale structure (LSS) formation, an {\it accretion} shock is identified as a stationary shock at about twice the virial radius of a cluster.  It is caused by a continuous accretion flow, typically along a filament of galaxies. An {\it infall} shock is caused by a dense clump (or group) of galaxies with its own intra-group medium penetrating the ICM 
in the early stages of a merging event. {\it Merger} shocks, as they have been called in the literature,  are outwardly propagating shocks that travel from the core of a merging event outward, typically illuminating the plasma (e.g., radio ``relics") on the outskirts of the ICM in the {\it later} stages of the merger.}  due to ongoing accretion, where the X-rays have been heated from their (non-detected) conditions in the more distant filament (e.g. En{\ss}lin et al. 1998; Bagchi et al. 2006).

To illuminate these alternatives, we re-looked at the existing data on the dynamics of mass accretion onto Coma (e.g., Adami et al. 2005). The average Coma radial velocity is 6925 km/s, and there is a well-known infalling group from the southwest associated with NGC~4839 (7362 km/s). In order to isolate the infalling galaxy velocity distribution, we plotted the SDSS spectroscopic velocities in a 2$^{\circ}$ wide band to the southwest of the Coma cluster  (Fig. \ref{vel_plot}).  We show a two component Gaussian fit to the velocity distribution, which includes both infall and virialized cluster periphery components.  The narrow, infall component, has a peak of 27 galaxies, a centre of 7450 km/s, and an rms dispersion of 300~km/s. The broader component, from the virialized galaxies on the cluster outskirts, has a peak of 30 galaxies, a centre of 7000 km/s, and an rms dispersion of 935~km/s. For comparison, Rines et al. (2003) find a mean velocity of 6973$\pm$45 and dispersion 957$\pm$30 in an analysis of the entire cluster neighbourhood including 1240 galaxies.  Fig. 
\ref{vel_plot} also shows a greyscale image of the smoothed galaxy surface-density (from SDSS spectroscopic data) for the range 6600 $<~v~<$ 8200 which best isolates the infalling galaxies.  In this velocity range we see a dramatic drop-off or ``wall" in the surface density of galaxies at the inner edge of the extended radio relic.  Remarkably, the transverse extent of the wall is 
also comparable to the relic, $\sim$2~Mpc.

\begin{figure}
\includegraphics[width=8cm]{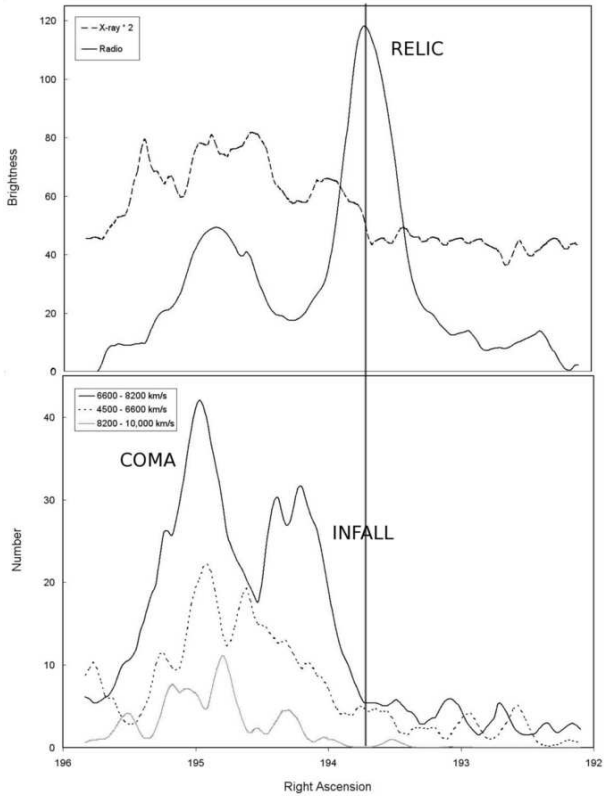}

\caption{\label{slice} Top: 1-D slice across the Coma halo and relic in 
radio and X-ray brightness; Bottom: Same slice except 2$^{\circ}$ wide 
through the optical surface density of SDSS galaxies in three velocity 
cuts. Note the drop in number of galaxies with 6600 $<$ v $<$ 8200 away from the cluster as the 
relic radio emission increases.}

\end{figure}

Fig. \ref{slice} shows a 1-D slice across the Coma halo and relic in 
radio (1.4~GHz) and X-ray\footnote{ The X-ray data were taken from an 
archival ROSAT PSPC image} brightnesses. There is no significant X-ray 
emission beyond the relic. We also plot the surface density of SDSS 
galaxies in three velocity bins along the same slice, with a width of 
2$^{\circ}$.  Moving away from the cluster, we find a sharp drop in the 
number of galaxies with 6600 $<$ v $<$ 8200 just as the relic radio 
emission increases. The other velocity bins, as expected, are dominated 
by the cluster itself and drop off gradually without any special 
behaviour at the relic position. Several infalling groups of galaxies 
have been identified near this region (e.g., Adami et al. 2005), but 
this is the first identification of a very broad transverse feature in the 
infall pattern into Coma. Rines et al. 2003 estimate a virial radius of 2.8~Mpc (using $\sim$1.3r$_{200}$, Eke, Cole 
\& Frenk 1996, and converting to H$_0$=70), so the galaxy wall is already 
within the virialized region but can still be isolated in space and velocity.

The correspondence between the wall of galaxies and the inner edge of 
the radio relic suggests a causal link. This poses a problem for the 
association between the radio relic and the NAT radio source NGC 4789, 
which is a possible source of seed relativistic electrons (En{\ss}lin et 
al. 1998). Given its radial velocity of 8365~km/s and the morphology of 
the bent jet, NGC 4789 is apparently on the back side of the cluster 
(moving away), while the wall of galaxies, if associated with the NGC 
4839 group, must be in front of the cluster (moving in). If the relic is 
indeed associated with the infalling wall of galaxies, then it would 
explain why the models of En{\ss}lin et al. (1998) underestimated its 
fractional polarization; we are viewing the relic almost edge on.  
However, similar to En{\ss}lin et al. (1998), we claim that the 
correspondence between the radio relic and the infalling wall of 
galaxies is evidence for this being a true large-scale infall 
shock and not an outwardly propagating merger shock.

 The polarization properties of the extended relic (Fig. \ref{gbt_pol}), namely that only the outside edge of the relic is polarized, is consistent with recent numerical simulations showing that relic emission consists of two regions (Paul et al. 2010). At the shock location, first-order Fermi acceleration dominates (and is polarized due to shock compression/alignment of the magnetic fields). The post-shock region is increasingly dominated by second-order Fermi acceleration from MHD turbulence, and is likely depolarized due to small-scale tangling of the field. High-frequency and resolution spectro-polarimetric observations of sufficient depth to detect the extended relic and solve for Faraday rotation are needed in order to confirm this paradigm. 

These optical and polarization data support the identification of the extended Coma relic as the only Mpc scale infall shock currently known. However, further investigation, especially with numerical simulations, is needed in order test the plausibility of this claim.

\subsection{Halo} The sharp synchrotron front seen in Figure \ref{wsrt_mos} is unique among classical GRHs. We will first compare the front to published and archival X-ray information, then examine possible origins for the synchrotron emission. Finally, we will examine the global radio vs. X-ray correlation of the radio halo in the context of GRH origin models.

Many deep X-ray observations have been taken of the Coma cluster with telescopes such as ROSAT, XMM, Chandra and Suzaku. X-ray soft (Bonamente et al. 2009) and hard excesses (e.g., Eckert et al. 2008) have been claimed in the Western infall region, which have been attributed to non-thermal IC emission. The non-thermal nature of the excess has been ruled-out/challenged by Wik et al. (2009) using Suzaku, though they confirm higher temperatures in the western region just interior to the synchrotron front. Neumann et al. (2003) also found an excess above a beta-model in the west, just interior to the synchrotron front. The exact kinematics of this western X-ray structure is not clear. Adami et al. (2005) argue that the northern and southern portions of the structure have different histories, with galaxy groups G12 and G14 associated with the south part of the structure, and appearing to be infalling from the rear.  

\begin{figure*}
\includegraphics[width=16cm]{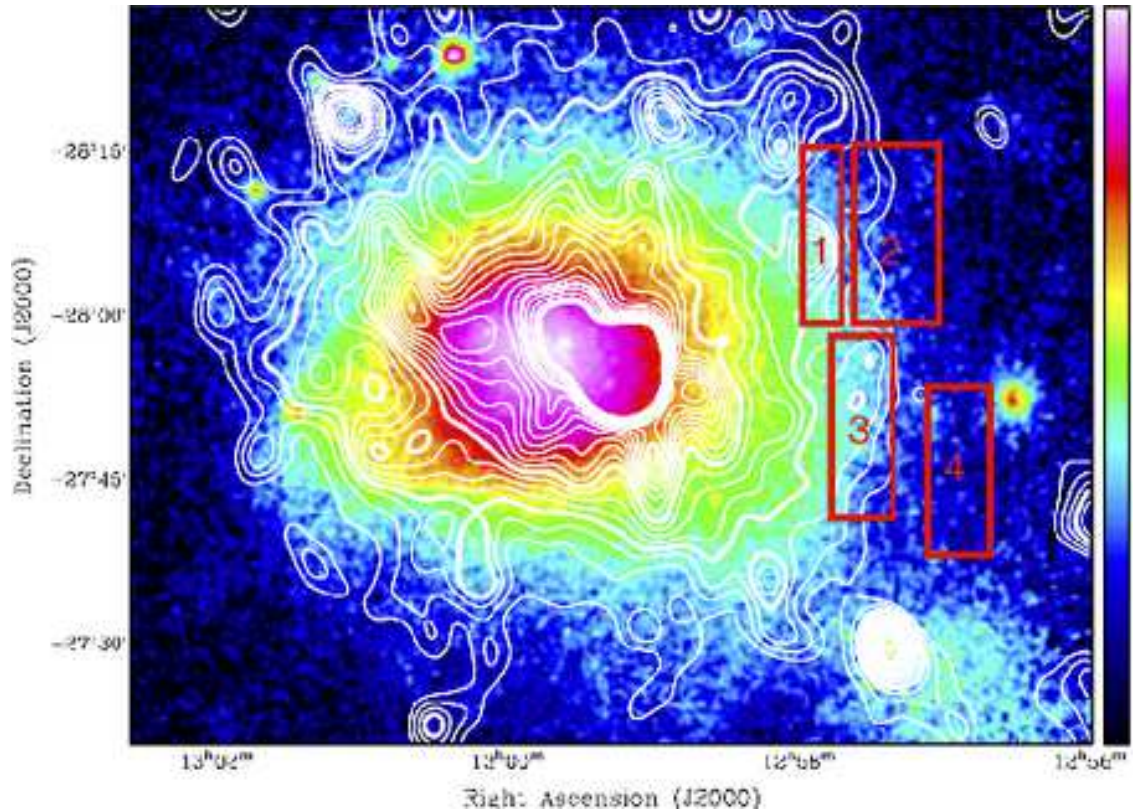}

\caption{\label{xray_diff} Contours are diffuse WSRT radio emission same over archival ROSAT PSPC diffuse X-ray emission. Radio contours start at 5 mJy~beam$^{-1}$ and increase in steps of 5 mJy~beam$^{-1}$ (beam of $4^{\prime} \times 3^{\prime}$). Boxes indicate regions where X-MM temperature fits where taken (see text).}

\end{figure*}

 Fig. \ref{xray_diff} shows archival ROSAT PSPC observations with our point-source subtracted WSRT contour image overlaid ($4^{\prime}\times 3^{\prime}$ resolution). There is a striking alignment of an apparent X-ray edge (Markevich et al. in preparation) in the ROSAT image with the synchrotron edge in the Southern portion of the front. The correlation in the South makes it likely that a shock front is responsible for both features, and the relativistic electrons were produced by shock-(re)acceleration.  The correlation is lost in the Northern region, where the synchrotron extends farther than the diffuse X-rays.   

A deep XMM-Newton mosaic of the Coma cluster (Schuecker et al. 2004; Wik et al. 2009) yielded a spatially resolved temperature map. Higher signal/noise fits were taken of the boxed regions indicated in Fig. \ref{xray_diff} (Alexis Finoguenov, private communication) which found [6.8$\pm$0.6, 16.3$\pm$2.9, 5.1$\pm$0.24, 2.46$\pm$0.21]~keV for boxes [1, 2, 3, 4], respectively. The factor of 2 increase in temperature across the Southern front toward the cluster is further evidence for the presence of a shock in this region. The Northern region, however, shows the opposite sign, with the temperature being much higher {\it outside} the synchrotron front. This, coupled with the poor correspondence between the synchrotron and X-ray surface-brightness profiles in the North, means that either the Western front is not a single shock structure or there are projection effects operating in the Northern region. Given the complex three-dimensional dynamics of the cluster, it is quite possible for part of the synchrotron front (in this case, the North) to be confused by unrelated X-ray brightness and temperature effects.

The general rise in temperature towards the northwest is most easily explained by shock-heating, either from an infall/accretion shock or an outward-propagating shock.  Outward propagating shocks can arise from merger events, or they could be currently driven by a subcluster/group leaving the cluster after passing through the core. The alternative to shock heating at the cluster periphery would be that the material was pre-heated up to 10-15~keV in the infalling filamentary structure before accreting onto the cluster.  Gas in filaments is normally in the range of 10$^{5}$ $<$ T $<$ 10$^{7}$~K, that is, a temperature less than 1~keV. It is implausible for gas in filaments to be heated to 10 keV or higher by structure formation shocks (e.g., Ryu \& Kang 2009). Those temperatures could be reached by AGN jet heating; however, any such heating to 10~keV or higher should be on smaller scales (Sijacki \& Springle 2006).

Of the various shock-heating alternatives, we somewhat favour the outward propagating shock driven by a group(s) leaving the cluster, because similar structures are commonly seen in numerical simulations (E. Hallman, private communication).  In these situations, the exiting group(s) set up a cold front along with a shock, and may produce only small X-ray enhancements, compared to the overall cluster emission, as is true for Coma.  We mention this primarily as an impetus for further study; simulations that produce the observed Coma temperature profile as well as a sharp front in the synchrotron material are needed to isolate the nature of the front and the overall dynamics of the situation.

Though not as dramatic as the X-ray/radio front presented here, possible shock fronts are being discovered in other clusters of galaxies through coincident X-ray/radio discontinuities (e.g., A520, A521, A754, A2744, and the Bullet cluster, Markevich et al. 2002, 2005; Krivonos et al. 2003; Henry et al. 2004; Markevich et al. in preparation). 
 
 High resolution Sunyaev-Zel'dovich effect (SZE) measurements (e.g., Mason et al. 2010) of this region would probe the pressure transition across the front, verifying the presence of a shock (Pfrommer et al. 2005) . Detection of polarized synchrotron emission with a position angle perpendicular to the front orientation would also support shock compression, though we might expect that any sharp transition in the synchrotron emitting plasma would also align the magnetic fields at the boundary region. 
 
\subsection{The Radio Cloud}
 In the discussion below, we assume that the 135' low surface brightness feature reported by Kronberg et al. (2007) and confirmed by us here (Fig. \ref{cloud}) is actually associated with the Coma cluster, and not a foreground galactic feature.  
This $\sim$4~Mpc ``radio cloud" is then the most extensive complex of synchrotron emitting plasma yet seen in cosmic large scale structure, and its presence has implications for the uncertain physics of cosmic-rays in those environments. Lifetimes for cosmic-ray electrons/positrons (CRe$^{\pm}$) radiating at 1.4~GHz reach a maximum of $\sim$10$^{8}$~yr  (Sarazin 1995, 1999), much shorter than the diffusion timescale within the cloud. This well-known result means that there must be in situ acceleration of CRe$^{\pm}$ working throughout the cloud volume. The relevant issues are: 1) What is the energy source? 2) What mechanism is accelerating or re-accelerating the CRe$^{\pm}$? 3) Where are the ``seed" e$^{\pm}$ coming from? We briefly address each of these.\\

\noindent 1) Gravitational infall, which we have argued in  $\S$4.1 \& $\S$4.2 is active on Mpc scales in the Coma cluster, is the most likely source of energization. This energy can be dissipated through shocks and/or turbulent heating (see 2). The $\sim$10$^{59}$-10$^{60}$~ergs of energy in the cloud (Kronberg et al. 2007) can easily be provided by the gravitational potential of the cluster. Similar energies can be provided by collective AGN or supernovae, though there will still need to be in situ (re)-acceleration of the CRe$^{\pm}$ (see 2 \& 3 below).\\

\noindent 2) There are three possible mechanisms for the (re)-acceleration of the emitting CRe$^{\pm}$ within the cloud; diffusive shock acceleration at infall and merger shocks (e.g., En{\ss}lin et al. 1998), turbulent re-acceleration of pre-relativistic plasmas (e.g., Brunetti et al. 2001), or secondary acceleration due to CRe$^{\pm}$ produced during CRp-p collisions (``hadronic" model).

\noindent 3) Seed e${\pm}$ can either come directly from the thermal plasma, relic relativistic plasma that was pre accelerated by AGN or during LSS formation, or as secondary products of CRp-p collisions. 

Observational constraints on these theories for an individual object such as the 4~Mpc Coma cloud are difficult to make. Hadronic models predict a strong correlation between X-ray and radio surface-brightness, which we show in $\S$4.4 is not the case for the Coma cluster itself. Diffusive shock acceleration should be differentiated from turbulent processes by the presence of polarization (do to magnetic field alignment at the shocks). However, both the GBT and WSRT images are confused by patchy Galactic emission across the field of view. The WSRT data has the added problem of internal Faraday depolarization at low frequencies, where even the bright polarized relic 1253+275 was not detected in our data after Faraday Rotation-Measure Synthesis (Brentjens \& de Bruyn 2005). Higher sensitivity, frequency, and resolution (e.g., 1.4~GHz EVLA ) observations of the polarized intensity, coupled with deeper X-ray imaging of the extended ICM,, are needed in order to resolve these important issues.

\subsection{The Global Radio vs. X-ray Correlation} Giant radio halos are expected to be tightly coupled to the X-ray emitting thermal plasma in clusters, the dominant form of baryons.  The magnetic fields are anchored in and amplified by dynamics of the thermal medium, and the relativistic particles ultimately derive their energy from the thermal medium's shocks and turbulence.  Thus, GRH models predict a pixel-to-pixel correlation between the radio and X-ray brightnesses within clusters, with some scatter since, e.g., shear can locally amplify magnetic fields without changing the thermal density. The correlation is typically expressed as a power-law of the form

\begin{equation} F_{Radio} = a\left(F_{X}\right)^{b}, \end{equation}   

\noindent where $F_{Radio}$ and $F_{X}$ are the radio and X-ray flux surface brightnesses and {\it a} and {\it b} are constants.  Here we focus on {\it b}, which parametrizes particle acceleration models.

\begin{figure}
\includegraphics[width=8cm]{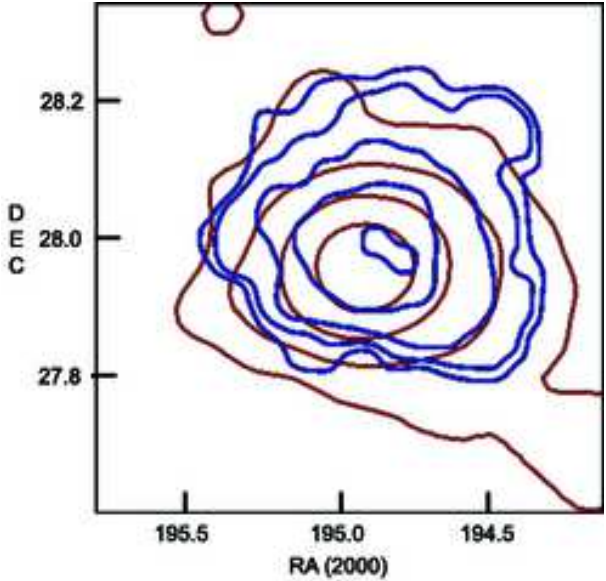}

\caption{\label{conts} Contours are WSRT radio (blue) and ROSAT X-ray 
(red) at 375$^{\prime\prime}$ resolution, emphasizing the difference in 
shape Contours are at factors of 2, starting at 75 counts (Xray) and 
0.001 mJy~beam$^{-1}$ (radio). Note that the presence of X-ray emission outside of the western front
blurs the sharp drop seen at full resolution.}

\end{figure}

Govoni et al. (2001) found a power law correlation with b=0.64$\pm$0.07, using 326~MHz WSRT data. We confirm their result using ROSAT PSPC data and our point-source filtered 352~MHz image (Fig. \ref{conts}), finding b=0.62$\pm$0.05. Govoni et al. (2001) illustrate, for primary electron models, that we actually expect b$\approx$1 if one assumes that the cluster is: 1) iso-thermal; 2) the magnetic, thermal, and cosmic-ray electron (CRe) energy densities are proportional to each other, with the same proportionality constants throughout the cluster (the proportionality premise); and 3) the radio spectral index is $\alpha$=1.

The $F_{Radio}$ vs. $F_{X}$ correlation found by Govoni et al. (2001) and the current work can also be compared to the unified model of diffuse cluster radio emission presented by Pfrommer et al. (2008), who find b=1.3-1.7 for secondary electron emission, and roughly linear (b=1, though with a broad distribution) for primary electron emission at lower radio surface-brightness levels.  The simulations of Pfrommer et al. (2008) assume proportionality between thermal and magnetic field energy densities when calculating synchrotron emissivity, which is likely why they get a roughly linear power-law for primary electron radio emission.  Pure secondary models also fail to match observations (e.g., Brunetti 2004; Donnert et al. 2009). Donnert et al. (2009) modeled the Coma cluster using a hadronic secondary model for the cosmic-ray populations and turbulence amplified magnetic fields seeded by galactic outflows. They found that the proportionality premise (except replace CRe with CRp for a secondary model) needed to be broken in order to fit the Govoni et al. (2001) data. In fact, the ratio of the CRp to thermal energy densities reached an unphysical $>$100\% at a cluster radius of $\sim$1~Mpc. In addition, the steepening of the halo's spectral index (Giovannini et al. 1993) at larger cluster radii could not be explained with a secondary model.

It is sufficient to look at the relative morphologies of the radio and X-ray halos (Fig. \ref{xray_diff} \& \ref{conts}) to see that any simple scaling relation will not be sufficient to capture the complex physical processes occurring throughout the ICM. It is likely that there are different physical processes creating the synchrotron emitting CRe in different regions of the radio halo; the X-ray emission also has significant contributions from both a relaxed-appearing $\beta-model$ component as well as residuals including the higher temperature region in the west (Neumann et al. 2009; Watanabe et al. 1999).  Turbulent re-acceleration models (Brunetti et al. 2001; Petrosian 2001) are able to produce extended GRH profiles such as the one seen in Coma (e.g., Brunetti \& Lazarian 2010), assuming a broadly distributed population of seed CRe$^{\pm}$. Our results in $\S$3.2 and $\S$4.2, which attribute some of the emission at large cluster radii to shock accelerated CRe$^{\pm}$, do not conflict with these results. In fact, the radial profile of the halo emission would be easier for any of the current models to fit if one's interest is only to model the central, x-ray tracking $\beta-model$ component.  Clearly the current models do not yet provide sufficient degrees of freedom to account for all of these various components or to predict any aggregate power-law correlation that might result. Further numerical and observational investigations are needed in order to uncover the dominant physical processes in still-evolving GRHs such as Coma.

\begin{figure}
\includegraphics[width=8cm]{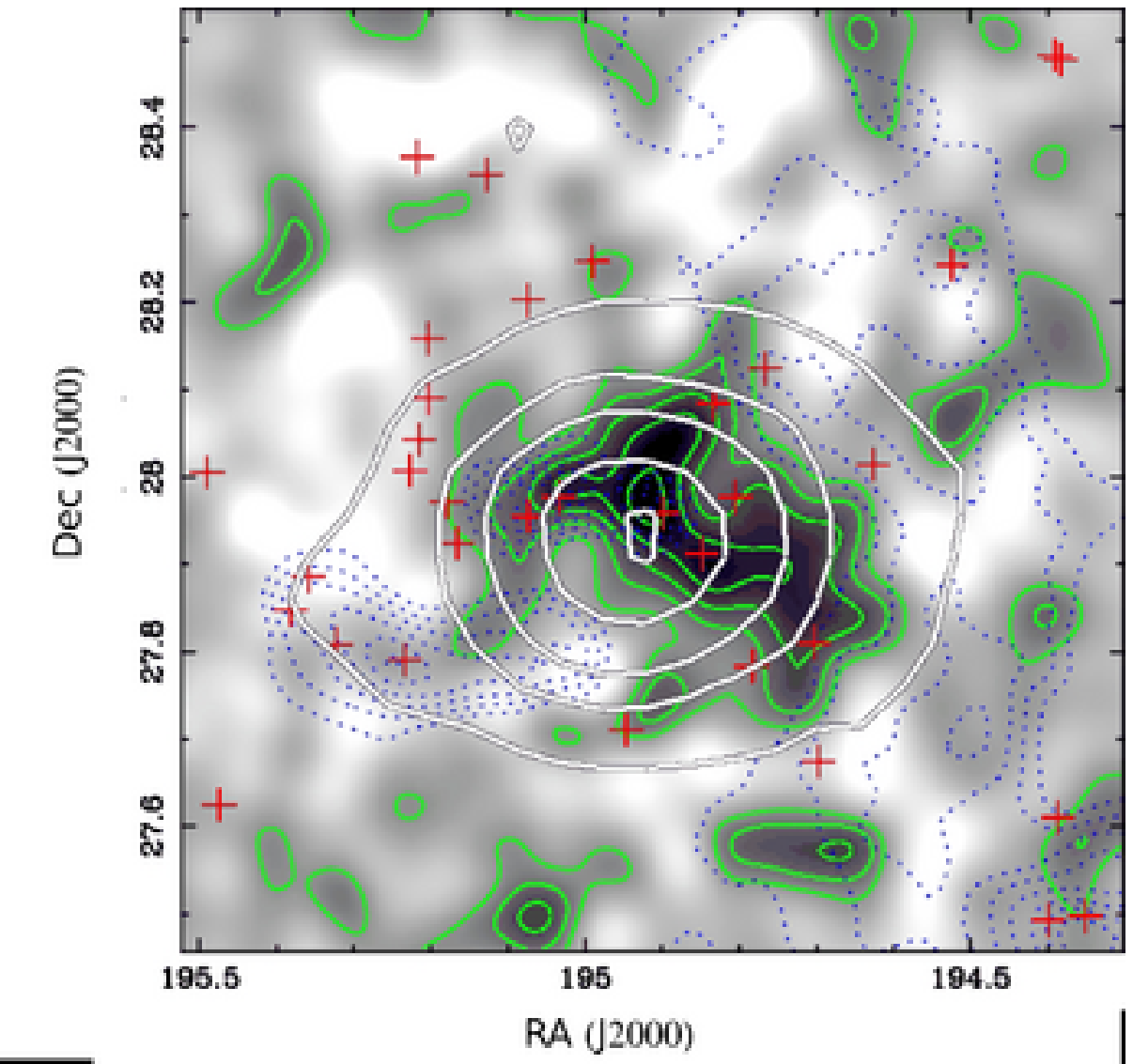}
\includegraphics[width=8cm]{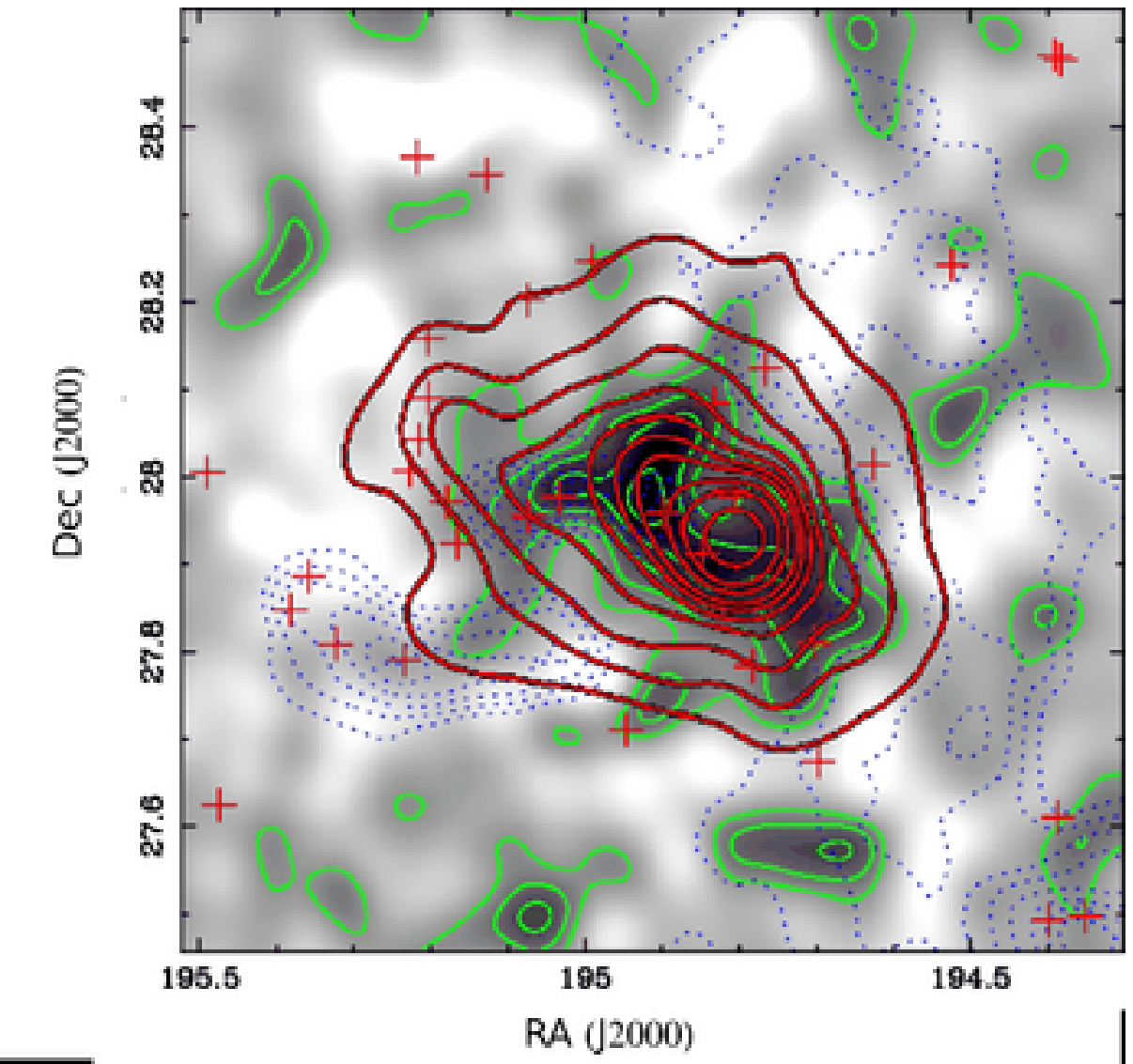}

\caption{\label{massXR}  Convergence map from weak lensing, showing the derived surface mass density in the centre of the Coma cluster, in greyscale and green contours, from Gavazzi et al. 2009.  Top: White contours of ROSAT broadband emission smoothed to 300".  Bottom: Radio contours of diffuse emission after filtering out point sources and convolving to 300", at levels of 0.05, 0.075, 0.1,... 0.2, 0.25, 0.3, 0.4 mJy/(300" beam). Dotted blue contours indicate the X-ray excess above a $\beta$-model, from Neumann et al. (2003).} \end{figure}

Figure \ref{massXR} compares the inner portion of the X-ray and diffuse radio emission with the mass surface density map of Gavazzi et al. (2009), derived from weak lensing.  They noted that the mass distribution did not correspond to the X-ray structure or to the distribution of bright galaxies.  Adami et al. (2009) did a very deep spectroscopic survey of these inner regions, and conclude that there are a number of background groups, including a major concentration (BMG) at z=0.054.  They suggest that  background structures may be responsible for some of the lensing-derived surface mass distribution.

 However, we now add the curious result that the brighter parts of the synchrotron halo are similar in size, shape and position to the derived surface mass distribution.  Residual diffuse emission from the cluster tailed radio galaxy NGC 4869 contributes only partially to this alignment at the brightest contours in Fig. \ref{massXR}. There does not appear to be a good physical reason for the synchrotron halo to track the mass of the cluster in this particular situation.  The X-ray emitting gas is offset from the apparent mass centroid, and we would expect the relativistic plasma to be well-coupled to that thermal gas, even given the brightness variations discussed earlier. One possible connection between the radio brightness and mass is that dark matter annihilation is expected to produce synchrotron radiation from its subsequent $\gamma \rightarrow ~ e^+e^-$ decay (Colafranceso et al. 2010).  Jeltema, Kahayias \& Profumo (2009) note that if such annihilation were responsible for a significant fraction of Coma's halo, it would be easily detectable in FERMI $\gamma$-rays.  The initial, 9 and 18 month results from FERMI do not yet show a detectable signal from Coma (Bechtol 2009; Ackermann et al. 2010).

An additional curious feature is that the derived surface mass distribution is surrounded quite closely, but does not overlap with, the western excess X-ray emission structure from Neumann et al. 2003 (see blue dashed contours in Figure \ref{massXR}). We tentatively conclude that most of the lensing mass distribution must be intrinsic to the Coma cluster, with the puzzle of why it appears so closely related to the synchrotron halo and the western X-ray excess is still unsolved.

\section{Summary} We conducted observations of Coma cluster at 1.41~GHz with the GBT and 352~MHz with the WSRT. Our key findings are:

\noindent$\bullet$ We detect an extension to the radio relic source 1253+275, making it $\sim$2~h$_{70}^{-1}$~Mpc in transverse extent.

\noindent$\bullet$ The extended relic lies just outside a previously unidentified $\sim$2~h$_{70}^{-1}$~Mpc wide ``wall" of infalling galaxies, of which the well known infalling group associated with NGC 4839 is a part. The physical relationship between this wall of galaxies and the radio relic is still unclear, but it argues for an infall shock origin for 1253+275. 

\noindent$\bullet$ There is a sharp, low surface-brightness ``front" of synchrotron emission on the Western edge of the 352~MHz radio halo where the brightness changes by 5~mJy~beam$^{-1}$. Its discrete edge, along with a corresponding edge in the X-ray surface-brightness and jump in temperature in the Southern portion of the front, suggests the presence of a shock or compression  on the western edge of the halo. 

\noindent$\bullet$ We confirm the existence of very low brightness emission enveloping the halo and relic, first reported by Kronberg et al. (2007) at 408~MHz.

\noindent$\bullet$ We find $F_{R}\propto F_{X}^{0.62\pm.05}$, in agreement with Govoni et al. (2001),  much flatter than primary and secondary models for the origin of the Coma cluster. We emphasize that a single scaling relationship does not capture the multiple components likely responsible for the different shapes and centres for the radio and X-ray emission.

\noindent$\bullet$ We find a correspondence between the brighter parts of the diffuse radio emission and the surface mass density distribution derived from weak lensing experiments.

We gratefully acknowledge help and advice from C. Adami, G. Brunetti, K. Dolag, P. Edmon, T. En{\ss}lin, E. Hallman, T. W. Jones, P. Kronberg, F. 
Miniati, C. Pfrommer, and D. Ryu. We thank A. Finoguenov for kindly providing the temperature fits across the western 
front, and M. Markevich for advice regarding the X-ray edge aligned with the Western front. We would also like to thank the anonymous referee for their careful readings and comments. 

Partial support for this work at the University of Minnesota comes from 
the U.S. National Science Foundation grant AST~0908688 and the NRAO GBT student
support program for SB. The Westerbork Synthesis Radio Telescope is operated by the 
ASTRON (Netherlands Foundation for Research in Astronomy) with support 
from the Netherlands Foundation for Scientific Research (NWO).  We acknowledge 
the use of NASA's SkyView facility located at NASA Goddard Space Flight 
Center. The Green Bank Telescope is operated by the National Radio
Astronomy Observatory (NRAO).  The NRAO
is a facility of the National Science Foundation operated under cooperative agreement by Associated Universities, Inc.
Digitized Sky Surveys were produced at the Space Telescope Science 
Institute under U.S. Government grant NAG W-2166. Archival observations 
obtained from XMM-Newton, an ESA science mission with instruments and 
contributions directly funded by ESA Member States and NASA, and ROSAT 
archives from HEASARC.

\end{document}